\documentstyle[11pt,aaspp4]{article}
\newcommand{\Ha}{H$\alpha$}
\newcommand{\Rv}{$R_V$}
\newcommand{\av}{$a_V$}
\newcommand{\ab}{$a_B$}
\newcommand{\ar}{$a_R$}

\newcommand{\HI}{\ion{H}{1}}

\newcommand{\Nh}{$N_H$}
\newcommand{\NII}{[\ion{N}{2}]}

\begin{document}

\title{Extraplanar Dust in the Edge-On Spiral NGC 891\footnote{ Based
on observations obtained by the WIYN Observatory which is a joint
facility of the University of Wisconsin-Madison, Indiana University,
Yale University, and the National Optical Astronomy Observatories.
Supplemental data are also provided by the NASA/ESA Hubble Space
Telescope, obtained from the data archive at the Space Telescope
Science Institute. STScI is operated by the Association of
Universities for Research in Astronomy, Inc. under the NASA contract
NAS 5-26555. }}

\author{J. Christopher Howk\footnote{howk@uwast.astro.wisc.edu} \&
     Blair D. Savage\footnote{savage@uwast.astro.wisc.edu}}
\affil{Department of Astronomy, University of Wisconsin-Madison
     \\ Madison, Wi. 53706}

\authoremail{howk@uwast.astro.wisc.edu}

% \begin{center}
% {\bf DRAFT: \today}
% \end{center}

\begin{abstract}
We present high-resolution (0.6\arcsec -0.65\arcsec) optical
broad-band images of the edge-on Sb galaxy NGC 891 obtained with the
WIYN 3.5-m telescope.  These BVR images reveal a complex network of
hundreds of dust absorbing structures far from the mid-plane of the
galaxy. The dust structures have a wide range of morphologies
and are clearly visible to $|z| \lesssim 1.5$ kpc from the mid-plane.
In this paper we discuss the general characteristics of the population
of absorbing structures, as well as physical properties of 12
individual features. These 12 structures are characterised by $N_H
\gtrsim 10^{21}$ cm$^{-2}$, with masses estimated to be more than $2
\times 10^5 - 5 \times 10^6$ M$_\odot$, assuming Galactic gas-to-dust
relationships.  The gravitational potential energies of the individual
dust structures, given their observed heights and derived masses, lie
in the range of $20-200 \times 10^{51}$ ergs, which represents the
energy input of at least tens to hundreds of supernovae.  Rough number
counts of extraplanar dust features find $\gtrsim$120 structures at
$|z| > 400$ pc with apparent B-band extinction $\gtrsim 0.25$ mag.  If
these have similar properties to those structures studied in detail,
the mass of high-$z$ gas associated with extraplanar dust in NGC 891
likely exceeds $2 \times 10^8$ M$_\odot$, which is $\sim$2\% of the
total neutral ISM mass of the galaxy.  The absorbing properties of
extraplanar dust structures in NGC 891 are best fit with $R_V \equiv
A_V / E(B-V) = 3.6 \pm0.4$.  A comparison of the new WIYN images with
an archival Hubble Space Telescope image of the central region of NGC
891 suggests that the quantities we measure for the extraplanar dust
features are not seriously affected by atmospheric blurring in the
WIYN images.

Some of the dust features seen in NGC 891 suggest supernova-driven
galactic fountain or chimney phenomena are responsible for their
production and are clearly associated with ionized gas structures
thought to be tracing the violent disk-halo interface of this galaxy.
However, other structures are not so readily associated with these
energetic processes.  We discuss several mechanisms which may produce
high-$z$ dust structures such as those seen in the WIYN and HST images
presented here.  It is, however, less than clear which of these
mechanisms are primarily responsible for the extensive extraplanar
dust structures seen in our images.  The data presented are part of a
larger program to search for and characterize off-plane dust
structures in edge-on systems.

\end{abstract}
 
\section{Introduction}

 The processes thought to eject matter away from the thin disks of
spiral galaxies will act upon both gas {\em and} dust.  What is the
fate of dust grains participating in various types of ejection
processes?  Are the ejection processes violent enough to completely
destroy the grains or is the destruction only partial?  Conversely,
does the presence of dust in galaxy halos provide important
information about the nature of the ejection processes?  Are the heavy
elements in halo matter mostly found in dust grain cores and thus
unavailable for important atomic radiative processes?  Does the
pressure of radiation acting on interstellar dust help support
material in galaxy halos?  Do ``galactic chimneys'' expel clean gas or
gas containing dust?  Very few of these questions can currently be
answered because very little is known about the existence of dust in
galaxy halos.

Though interstellar dust grains may play a significant role in the
evolution of material undergoing circulation between the disks and
halos of spiral galaxies, this role has not received much attention
from an observational point of view, and is only beginning to be
addressed theoretically.  Sofue and his coworkers have suggested that
dust filaments they identify in inclined galaxies may be the results
of magnetic phenomena or at least may be tracing the magnetic fields
in these galaxies (Sofue 1987; Sofue et al. 1994).  The Sofue et
al. (1994) paper discusses the appearence of the dust features in NGC
253 and suggests various models for the range of dust structure
morphologies observed.  They argue that the loops and filaments seen
in this galaxy are associated with magnetic fields and may be caused
by ``magnetic inflation'' or by radiatively driving dust along
magnetic field lines.  Unfortunately, the true $z$-extents of the dust
features identified by Sofue et al. (1994) in NGC 253 are rather
uncertain because the galaxy is inclined $\sim 12^\circ$ from edge-on
orientation.

In an interesting series of theoretical papers, Ferrara et al. (1991),
Franco et al. (1991) and Ferrara (1993) have recently re-investigated
the effects of radiation pressure on dust grains, including the
possibility of radiatively driving dust grains to large $z$-distances
in spiral galaxies.  Their work suggests that radiation pressure on
grains can have a significant effect on the kinematics of low-halo
clouds of material and may be the dominant factor in determining the
dust distribution in hot halos of galaxies.  Though these workers have
begun to address the problem, it is still not clear, theoretically or
observationally, what role dust plays in the large-scale circulations
that may occur between the disks and low-halos of spiral galaxies.

With this paper we begin a project to study off-plane dust structures
in highly-inclined galaxies, the goal of which is to learn more about
the interstellar disk-halo interface in galaxies by studying
extraplanar interstellar dust.  In this first paper we present images
of NGC 891 taken with the 3.5-m WIYN telescope under excellent seeing
conditions (0.6-0.65\arcsec\ FWHM).  The images show a wealth of dust
structures beyond the thin disk of the galaxy, some stretching to $|z|
\gtrsim 1.5$ kpc.  Several of these absorbing structures have been
mentioned by previous workers, though no in-depth discussion of their
properties or discussion of their role in the disk-halo interface
exists.  While in many ways NGC 891 is not typical of all galaxies,
the images presented herein unequivocally state the case for the
presence of dust beyond the thin disk of at least one spiral galaxy.

Sect. \ref{sec:n891} gives a brief summary of previous observations of
extraplanar interstellar material in NGC 891.  Sect. \ref{sec:obsred}
describes the observations and our reduction methods; the properties
of extraplanar dust in NGC 891 are discussed in Sect. \ref{sec:dust},
including a discussion of individual features and an analysis of the
absorbing properties of the extraplanar dust structures.
Sect. \ref{sec:hst} compares the WIYN images with archival Hubble
Space Telescope (HST) images to assess the degree to which seeing
affects our conclusions.  Sect. \ref{sec:discussion} is a discussion
of the possible mechanisms for populating the halo of spiral galaxies
with dusty material, and Sect. \ref{sec:summary} summarizes our
principal results.

\section{Extraplanar Gas in NGC 891}
\label{sec:n891}

NGC 891 is a well studied Sb galaxy seen near edge-on ($i\geq
88.6^\circ$, Rupen 1991).  It has been extensively observed over most
of the electromagnetic spectrum and has been particularly important
for the study of extraplanar gas.  The properties of the diffuse
extraplanar H$\alpha$ emission in this galaxy have been discussed by
numerous workers (Dettmar 1990; Rand et al. 1990; Keppel et al.  1991;
Pildis et al. 1994a).  These studies find diffuse ionized gas (DIG)
extending to at least 40\arcsec\ from the midplane of the galaxy, or
$|z|$=2 kpc assuming a distance of 9.5 Mpc (van der Kruit \& Searle
1981; Keppel et al. 1991), with some studies finding evidence for DIG
extending further than $|z|$=5.5 kpc (Rand 1997).  The recent Rand
(1997) study suggests the ionizing spectrum responsible for this gas
is significantly harder than that of the local Milky Way thick disk
DIG, based upon a comparison of the \ion{He}{1}/\Ha\ line ratio in
this gas to the upper limit for Milky Way gas derived by Reynolds \&
Tufte (1995).  The inferred gas temperature from his observations is
$T<$10,000-13,000 K.

Radio continuum studies of NGC 891 have similarly revealed an extended
cosmic-ray halo which is generally correlated with the H$\alpha$
emission (Dahlem, Dettmar \& Hummel 1994).  The \Ha\ and radio
continuum observations suggest a position dependent scale height of
$\sim0.6-1.3$ kpc [Rand (1997) also finds a second component with
scale height $\sim$5-6 kpc] and show a strong asymmetry, with the NE
side of the galaxy having enhanced emission (Dahlem et al. 1994).  The
radio data show significant polarization, with the orientation of the
polarization vectors suggesting the large-scale magnetic field
structure is parallel to the plane through most of the low-halo,
though there may be slight deviations toward more vertically oriented
fields further from the center of the galaxy (Dumke et al. 1995;
Sukumar \& Allen 1991).  The patterns of polarization in optical and
infrared polarization maps have been interpreted as evidence for
vertically oriented or non-toroidal magnetic fields through most of
the galaxy (Scarrott \& Draper 1996; Fendt et al. 1996; Jones 1997;
Wood \& Jones 1997).  The difference in the suggested orientations for
the magnetic field may, in part, be due to the differing resolutions
of the two data sets, or possibly to optical depth effects in the
visible/IR data.  Though the ionized halo of NGC 891 is thick enough
to provide significant Faraday rotation at longer radio wavelengths,
the observations by Dumke et al. (1995) and Sukumar \& Allen (1991) at
2.8 and 6 cm, respectively, should be relatively free from the effects
of Faraday rotation and provide an accurate measure of the orientation
of the magnetic field.  Both these studies conclude that the
orientation of the field lines is predominantly parallel to the plane
of the galaxy.

NGC 891 has been observed in CO by several workers
(Garc\'{\i}a-Burillo et al. 1992; Handa et al. 1992; Scoville et
al. 1993).  Garc\'{\i}a-Burillo et al. (1992) found an extended ($|z|
\lesssim$1.1-1.5 kpc) thick disk or ``halo'' of CO emission within
$\pm3.5$ kpc of the galaxy center using single-dish observations.
They argue that this component cannot be due to a warping or flaring
of the disk based upon the velocity structure of the extended-$z$
material.  Indeed, Handa et al.  (1992) also find evidence for CO away
from the midplane in the form of a ``molecular spur'' that stretches
$\sim$520 pc into the halo; they do not discuss the existence of an
extended thick disk.  Scoville et al. (1993) have presented
aperture-synthesis data for the CO $J$=1-0 transition.  They find no
evidence for the extended component of Garc\'{\i}a-Burillo et al.,
though their data are only sensitive to variations on spatial scales
$\lesssim$30\arcsec ($\lesssim$1.4 kpc).  The extended component may
show evidence for a higher CO ($J$=2-1)/($J$=1-0) ratio than the disk
material (Garc\'{\i}a-Burillo et al. 1992).  If this is indeed true,
the higher-$z$ material may be warmer than the disk gas; Scoville et
al. point out that their observations, made only in the $J$=1-0
transition, could possibly have missed an extended warm component.

Both the CO and \HI\ (Rupen 1991; Sancisi \& Allen 1979) observations
suggest that NGC 891 lacks an appreciable warp.  The deviations from
the plane in \HI\ are less than 3\%\ of the radius (Sancisi \& Allen
1979).  Sancisi \& Allen, however, argue that the \HI\ disk flares at
large distances.  They derive a FWHM thickness for the \HI\ layer of
$< 1$ kpc for the inner regions of NGC 891 and $\sim$1-2 kpc for
radial distances $R > 20$ kpc from the center of the galaxy.  A flared
disk would not explain the extended CO component observed by
Garc\'{\i}a-Burillo et al. (see their Sect. 3.2), though Scoville et
al. (1993) observe a slight increase in the CO $z$-extent with
projected distance from the galaxy's center.  Recently, Swaters et
al. (1995, 1997) have reported on new, very sensitive \HI\
observations of NGC 891.  The new data, and their modelling of these
data, suggest that a more appropriate model for the velocity structure
and vertical distribution of \HI\ in NGC 891 is a composite model
having both a thin and a thick disk, with the latter lagging the
former by $\sim$25 km s$^{-1}$ (Swaters et al. 1997; see also van der
Hulst 1996).  These new data also exhibit \HI\ extending to $| z |
\lesssim 5$ kpc and may show enhanced $z$-structure in the regions of
the disk that have enhanced \Ha\ and radio continuum emission (van der
Hulst 1996).

Bregman \& Pildis (1994) and more recently Bregman \& Houck (1997)
have observed a diffuse X-ray halo surrounding NGC 891 with the ROSAT
observatory.  This halo has a vertical surface brightness HWHM of
$\sim$2 kpc; fits to the data suggest a vertical gaussian density
scale-height of $\sim$3.5 kpc and T$\sim3.5 \times 10^6$ K (Bregman \&
Houck 1997).  They calculate the mass of the X-ray emitting gas to be
$\sim 4 \times 10^7$ M$_\odot$; the mid-plane density and
pressure best-fit values are 0.026 cm$^{-3}$ and 10$^5$ K cm$^{-3}$,
respectively (for filling factors of unity; see Bregman \& Houck
1997).  Much of the observed emission from the halo of NGC 891 has
been interpreted (Bregman \& Pildis 1994; Pildis et al. 1994a; Rand et
al. 1990; Dettmar 1990) as being at least consistent with the galactic
fountain model of disk-halo circulation (Shapiro \& Field 1976;
Bregman 1980) or the chimney-dominated model of Norman \& Ikeuchi
(1989).  If the hot material traced by the X-ray emission is indeed
participating in a fountain flow, Bregman \& Houck predict gas is
cooling at a rate of 0.09 M$_\odot$ yr$^{-1}$.

\section{Observations and Reduction}
\label{sec:obsred}

The primary observations presented herein were taken on the night of 3
December 1996 with the WIYN 3.5-m telescope at Kitt Peak National
Observatory.  The WIYN imager is a thinned 2048$^2$ STIS CCD with 21
$\mu$m pixels.  Placed at the f/6.5 Nasmyth focus of the telescope
each pixel corresponds to 0.196\arcsec\ on the sky, which is 9.0 pc at
NGC 891 using the adopted distance of 9.5 Mpc.  The images presented
here were taken under non-photometric conditions.  This should not
affect our analysis, as our measurements rely upon {\em relative}
photometry.  The seeing conditions during the observations were
excellent (0.6\arcsec -0.65\arcsec).

We obtained B,V and R broad-band images of NGC 891; Table
\ref{table:log} gives the properties of these images.  The resolution
given in this table refers to the seeing-limited resolution for the
WIYN images and is expressed as the FWHM of Gaussian profiles fit to
direct measures of stellar image sizes using the IRAF\footnote{IRAF is
distributed by the National Optical Astronomy Observatories, which is
operated by the Association for Research in Astronomy, Inc., under
cooperative agreement with the National Science Foundation.} routine
{\tt IMEXAM}.  The images have been bias-subtracted and flat
field-corrected in the usual manner within IRAF.  The flat fields were
derived from observations of the ``Great White Spot'' in the telescope
dome.  Registration of the images was done using a grid of stars
common to all of the images, and the rms errors in the alignment of
the three images are approximately 5-10\% of a pixel.  Similarly, we
have used several stars, whose coordinates were measured from
Digitized Sky Survey images, to solve for the astrometric plate
solution coefficients of our images.  The rms error in using the plate
solution to derive coordinates is approximately 0.7\arcsec.  The sky
background was determined from the average of numerous star-free
regions throughout the chip.

Part of our work makes use of supplementary observations from WIYN and
HST.  In Sect. \ref{subsec:individual} we discuss \Ha\ observations
taken with the WIYN telescope.  These images were taken through a
narrow-band interference filter ($\lambda_{cen} = 6570$ \AA; FWHM = 16
\AA), and used R-band images for continuum subtraction.  The exposure
time and resolution (after slight smoothing) of the final \Ha\ image
are given in Table \ref{table:log}.  We have also used an archival HST
snap-shot image of the central region of NGC 891 to supplement our
WIYN images.  The details of the reductions and analysis of this image
are given in Sect. \ref{sec:hst}.  The exposure time and resolution
for this image are also given in Table \ref{table:log}.

\begin{planotable}{ccccc}
\tablenum{1} 
\tablewidth{0pc}
\tablecolumns{5}
\tablecaption{Observing Log}
\tablehead{\colhead{Filter} & \colhead{Exposure} & \colhead{Pix. Scale} & 
\colhead{Resolution} & \colhead{Resolution}\\
\colhead{} & \colhead{(sec.)} & \colhead{(arcsec pix$^{-1}$)} & \colhead{(arcsec)}
& \colhead{(pc)}}
\startdata
\multicolumn{5}{c}{WIYN Images} \nl
B & 500 & 0.196 &  0.65 & 30 \nl
V & 120 &  0.196 & 0.65 & 30 \nl
R & 300 &  0.196 & 0.60 & 28 \nl
H$\alpha$ & 2100 & 0.196 & 0.90 & 41 \nl
 & & \nl
\multicolumn{5}{c}{HST WFPC2 Images} \nl
F606W & 2$\times$90 & 0.1 & 0.1\tablenotemark{a} & 5\tablenotemark{a} \nl
\enddata
\tablenotetext{a}{The HST PSF is undersampled in the WFC CCD.}
\end{planotable}

\section{Extraplanar Dust in NGC 891}
\label{sec:dust}
 
\subsection{General Comments}

Figure \ref{fig:WIYNcolor} shows a three-color image produced by
combining the B,V and R exposures\footnote{This image was prepared by
Dr. Nigel Sharp of NOAO.}.  The CCD was rotated to image the greatest
part of the galaxy, hence North lies 20$^\circ$ clockwise from the top
of this image with East being similarly offset from the right of the
image.  The color image has a slightly worse resolution than the
original images.  We show the B-band view of NGC 891 in Figure
\ref{fig:WIYNfull}.  North and East are labelled on this image, and
the physical scale is given by the bar in the lower left of the
figure.  One's eye is immediately drawn to the prevalent filaments of
absorbing dust, particularly seen against the light of the stellar
bulge.  Figure \ref{fig:WIYNcenter}(a) shows a close-up of the bulge
area from the B-band data, while Figure \ref{fig:WIYNne}(a) displays a
B-band close-up of the region imaged in \Ha\ by Pildis et al. (1994a)
to NE of the bulge, approximately 5 kpc projected distance from the
center of the galaxy.  Figure \ref{fig:WIYNsw}(a) shows the B-band
view of the disk to the SW of the central bulge.  Figures
\ref{fig:WIYNcenter}(b), \ref{fig:WIYNne}(b) and \ref{fig:WIYNsw}(b)
show unsharp-masked versions of the same regions, which were produced
by dividing the original image by a version smoothed with a gaussian
filter having FWHM=15 pixels (135 pc).  To aleviate problems of severe
over-subtraction, several stars were replaced with the median value of
the surrounding pixels in producing the unsharp-masked versions of
these images.  Regions in the unsharp-masked images where stars have
been removed, and hence regions that should be viewed as erroneous,
are identifiable by comparing the unsharp-masked versions of the
figures with the original B-band data.  In these figures the physical
scale, measured in the $z$ and R directions from the center of the
galaxy, is given in kpc along the edges; the scale given has the same
orientation as that defined by Garc\'{\i}a-Burillo et al. (1992).  A
number of the more prominent absorbing features are marked in Figures
\ref{fig:WIYNcenter}(b) and \ref{fig:WIYNne}(b) for further
discussion.  Table \ref{table:observed} contains the locations,
physical sizes, morphologies and extinction data for each of these
complexes; Table \ref{table:derived} contains the quantities derived
from these data (see Sect.  \ref{subsec:derived}).  We have chosen
these features because they unambiguously reveal dust far from the
midplane of this galaxy.  Practically this implies that we have
preferentially chosen relatively thick clouds that lie on the near
side of the galaxy.  Given the line-of-sight distance ambiguity, we
have no assurance that these features are indeed individual connected
structures.

\begin{planotable}{ccccccccccl}
\tablenum{2}
\tablewidth{0pc}
\tablecolumns{10}
\tablecaption{Observed Properties of Selected Dust Features}
\tablehead{ \colhead{Feature}   & \colhead{R.A.\tablenotemark{a}}    &
\colhead{Dec.\tablenotemark{a}} & \colhead{Length\tablenotemark{b}}  
&\colhead{Width\tablenotemark{c}}& \colhead{$z$\tablenotemark{d}}     
&\colhead{$a_{\mathrm B}$\tablenotemark{e}} 
&\colhead{$a_{\mathrm V}$\tablenotemark{e}} 
& \colhead{$a_{\mathrm R}$\tablenotemark{e}} 
&\colhead{Morphology\tablenotemark{f}}\\
        \colhead{} & \colhead{(B1950.0)} & \colhead{(B1950.0)} &
        \colhead{(pc)} & \colhead{(pc)} & \colhead{(pc)} &
        \colhead{} &\colhead{} & \colhead{} & \colhead{} }
\startdata
1 & 2$^h$\ 19$^m$\ 22.0$^s$ &  42$^\circ$ 06\arcmin\ 40\arcsec   
    & 360 & $\sim$300 & $\lesssim800$  & 1.55 & 1.31 & 1.05  & 
   Vert. column\nl
2 & 2 \ 19 \ 20.4 & 42 \ 06 \ 44 & 100 & 50  & 1450  & 0.84 & 0.65 &
      0.54  &  Elongated cloud\nl
3 & 2 \ 19 \ 22.6 &  42 \ 07 \ 21 & 210 & 90 & 1120 & 0.45 & 0.37 &
    0.31  & Irr. cloud\nl
4 &  2 \ 19 \ 23.8 &  42 \ 07 \ 36 & 450 & 75 & $\lesssim1350$ & 1.11 &
    0.80  & 0.65  & Vert. column\nl
5 &  2 \ 19 \ 25.6 &  42 \ 06 \ 49 & 145 & 45  & 950  & 0.59 & 0.49 &
    0.37  & Irr. cloud\nl
6 &  2 \ 19 \ 25.4 & 42 \ 06 \ 56 & 250 & 90 & 700 & 0.75 & 0.65 & 
    0.47  & Irr. cloud\nl
7 &  2 \ 19 \ 26.5 &  42 \ 06 \ 51 & 110 & 60  & 1350  & 0.37 & 0.30 &
    0.24 & Cometary cloud?\nl
8 &  2 \ 19 \ 26.1 & 42 \ 07 \ 23 & 360 & 70 & $\lesssim750$ & 0.59 &
    0.52 & 0.36 & Vert. column\nl
9 &  2 \ 19 \ 27.2 & 42 \ 07 \ 46 & 400  & 350 & $\lesssim 1000$ &
    0.61 & 0.50 &  0.40 & Vert. column   \nl
10 &  2 \ 19 \ 29.4 &  42 \ 08 \ 47 & 600 & 270 & $\lesssim900$  & 0.59 &
    0.52 & 0.36 & Vert. cone \nl
11a & 2 \ 19 \ 30.2 &  42 \ 09 \ 01 & 250 & 65  & $\lesssim750$ & 0.80 &
    0.59 & 0.50: & Shell wall \nl
11b & 2 \ 19 \ 30.7 &  42 \ 09 \ 13 & 250 & 55  & $\lesssim800$ & 0.40 &
   0.39 & 0.30: & Shell wall\nl
12 & 2 \ 19 \ 26.9 & 42 \ 09 \ 00 & 600: & 80 & $\lesssim$1000 &
    1.00 & 0.82 &  0.67 & Irr. vert. column \nl
\enddata
\tablenotetext{a}{Approximate coordinates of the dust feature.}
\tablenotetext{b}{Long dimension in parsecs.}  
\tablenotetext{c}{Short dimension in parsecs.}  
\tablenotetext{d}{Height above the midplane, or the limit to which very
extended features can be traced.} 
\tablenotetext{e}{Apparent extinction for the BVR wavebands in
magnitudes, as defined in the text.}  
\tablenotetext{f}{Morphological description; these are very subjective.}  
\end{planotable}

On close inspection one can identify features that seem to be rooted
in the disk and stretch primarily away from the midplane (e.g.,
features 4 and 10); other absorbing features seem to be unconnected to
the midplane dust lane (e.g., features 5 and 7).  It is also
interesting to note that the observed features are anything but
simple.  The appearence of the dust structures is mottled or
``chaotic,'' in the non-mathematical sense.  Much of the complexity
may be due to the superposition of several structures along the line
of sight, though some features seem quite well aligned (e.g., feature
4), suggesting coherence and physical association.  Further, while
some of the dusty filaments protruding from the disk are easily
associated with regions of star-formation in the disk, others are not.
The morphology of some features leads one to think of supernova-driven
activity: cones and shell-like structures (e.g., features 10 and 11
below).  Others are less suggestive in this respect (e.g., features 1
and 5).  There are features tracable to $|z| \gtrsim 1$ kpc throughout
the galaxy.  In the central region, where the distribution of
background light is more beneficial for probing high-$z$ material,
there are clouds that can easily be found 1.5 kpc from the midplane
(e.g., features 2 and 7).  There is further evidence, which is thus
far inconclusive, in color maps of the galaxy for material at least
1.7 kpc from the midplane and possibly beyond.  We cannot, at this
time, set an upper $z$-height to which the dusty material extends.

The lack of a clear asymmetry about the midplane of the galaxy reveals
that these structures cannot be attributed to a warp.  Given the
relatively certain association of some features with star-formation,
the evidence for foreground stellar light towards the features (see
Sect \ref{subsec:extinction}) and the new interpretation of the
thickened \HI\ distribution (Swaters et al. 1997), we attribute these
dust structures to inner galaxy ($R<20$ kpc) high-$z$ features rather
than structures associated with a flared gas layer at large radii ($R
> 20$ kpc). 

We note that Sofue (1987) and Sofue et al. (1994) identify vertical
absorbing features emanating from spiral arms in the inclined galaxies
NGC 253 and NGC 7331.  They argue that these features must stretch
vertically, given that they cross several spiral arms and seem to lie
mostly-perpendicular to the major axes of the galaxies.  The
photographs presented in these papers show structures (their
``vertical dust streamers,'' as well as loops and arcs) reaching
heights of $z\lesssim 1.5$ kpc from their bases, with widths of order
50-100 pc, quantities similar to those derived for the structures
listed in Table \ref{table:observed}.

The WIYN images of NGC 891 also show what appears to be a background
of poorly-defined absorbing structures, particularly against the bulge
of the galaxy.  This complex background is better seen in the
unsharp-masked versions of the B-band image, shown in Figures
\ref{fig:WIYNcenter}(b), \ref{fig:WIYNne}(b) and \ref{fig:WIYNsw}(b),
which clearly show the wealth of structure that is less obvious than
the most prominent features.  These less-distinct absorbing features
presumably represent a tangled superposition of many individual
high-$z$ features at varying distances through the thick disk of the
galaxy.  Some of these background absorbing features are
distinguishable as filamentary structures in an HST image of the bulge
(see Sect. \ref{sec:hst} below), though there appears to be a confused
web of background absorbers in the higher-resolution HST image as
well.

\subsection{Extinction Properties}
\label{subsec:extinction}

The extinction data for the features identified in Figures
\ref{fig:WIYNcenter} and \ref{fig:WIYNne} are given in Table
\ref{table:observed}.  The quantities \ab, \av\ and \ar\ are the
``apparent extinctions'' towards the dust features for the B, V and R
bandpasses, respectively (given in magnitudes).  We define the
apparent extinction $a_\lambda$ for a given waveband $\lambda$ as
\begin{equation}
        a_\lambda = -2.5 \log( S_{dc, \lambda} / S_{bg, \lambda} )
\label{eqn:extinct}
\end{equation}
where $S_{dc, \lambda}$ is the surface brightness measured towards a
dust cloud and $S_{bg, \lambda}$ is the surface brightness of the
local background. 

We have measured the appropriate surface brightness ratios for
individual dust features along five-pixel wide strips that pass
through the feature of interest.  Thus, each individual strip
represents an average of the data over 0.98\arcsec\ or 45 pc.
Typically measurements of 10-20 strips, each at a different position,
were taken for each dust structure.  Where possible the measurements
were taken both along strips parallel and perpendicular to the
galactic plane, roughly parallel to both the major and minor axes of a
dust feature, depending upon its orientation.

Measuring the ratio for an individual strip through a dust structure,
$(S_{dc, \lambda} / S_{bg, \lambda})_i$, involved defining a
background surface brightness, which was complicated by the multitude
of absorbers along the line of sight, and measuring an appropriate
on-cloud brightness.  The background brightness was derived by
interpolating between regions in the locale of the feature that seem
to be free of dust. For ease of measurement and to minimize
contamination caused by seeing effects, the surface brightness towards
the dust complex was chosen to be the point giving the minimum value
of $(S_{dc, \lambda} / S_{bg, \lambda})_i$.

The final ratio $S_{dc,\lambda} / S_{bg,\lambda}$ used to derive the
values of $a_\lambda$ given in Table \ref{table:observed} was the
average of all the individual measurements $(S_{dc, \lambda} / S_{bg,
\lambda})_i$ for a feature.  The values of $a_\lambda$ given in this
table are thus spatial averages of the individual features, since each
measurement $(S_{dc, \lambda} / S_{bg, \lambda})_i$ sampled a
different position through the structure.  While the observations were
taken in non-photometric conditions, the measurement of relative
surface brightnesses is still valid.  We have made no correction for
the presence of emission lines in our images.  The most significant
contaminator is likely \Ha\ in the R-band images.

The technique we have adopted to measure the ratio $S_{dc,\lambda} /
S_{bg,\lambda}$ may, in the individual measurements, tend to be
systematically influenced by noise features, i.e., the minimum of the
ratio is almost always in a negative noise feature, systematically
lowering the observed ratio.  This is not, in the end, particularly
problematic, for the photon noise that systematically affects the
measurements is generally the smallest addition to the mix of
uncertainties.  Each individual measurement is itself taken from the
average of five adjacent 0.196\arcsec\ pixels, and the apparent
extinction quoted for a given feature represents the average of many
positions along the absorbing dust structure.  The photon noise in the
individual measurements are certainly less than the uncertainty
associated with defining the continuum level, or $S_{bg,\lambda}$.
Further, the spatial fluctuations of the ratio $S_{dc,\lambda} /
S_{bg,\lambda}$ in the 10-20 measurements taken at different positions
along the absorbing feature tends to dominate the photon noise of the
individual measurements.  The standard deviation of the individual
measurements about the mean for a given feature is generally
$\sim$10\%.  The systematic affect of the photon noise is minor
compared to the other uncertainties involved in our measurement
scheme.

The values $a_\lambda$ of Table \ref{table:observed} do not
necessarily represent the true extinction, $A_\lambda$, which includes
contributions from absorption plus scattering, because of scattering
into the line of sight.  Further, $a_\lambda$ differs from the true
extinction in that there may be stars on the near side of these
features whose light is not absorbed by the dust.  It is possible,
even in light of these problems, to derive approximate properties of
the dust in NGC 891.  If we define $x$ as the fraction of galactic
light in a given waveband emitted in front of the dust features, we
may follow Gallagher \& Hunter (1981) and Knapen et al. (1991) by
writing
\begin{equation}
   S_{dc, \lambda} / S_{bg, \lambda} =
                x + (1-x)e^{- \tau_\lambda} .
\label{eqn:xtau}
\end{equation}
The quantity $\tau_\lambda$ is assumed to roughly approximate the
extinction optical depth through the cloud at wavelength $\lambda$,
i.e., $\tau_\lambda$ contains contributions from pure absorption plus
scattering out of the line of sight.  With this formulation it is
important to note that equation (\ref{eqn:xtau}) neglects the effects
of light scattered off of dust into the line of sight.  The scattering
phase function for interstellar dust is very forward-throwing.  This
implies that most of the light scattered from large angles to our
sightline will not enter our beam; the fraction that does is likely
balanced by scattering of background stellar light out of the beam,
implying that $\tau_\lambda$ is not precisely equivalent to the true
absorption plus scattering optical depth.  In our application of this
equation we will assume that the value of $x$ is not wavelength
dependent.  Equation (\ref{eqn:xtau}) also assumes the absorbing
features we identify are localized in space, i.e., not intermixed with
the stars, which is likely reasonable given the discrete appearance of
the extraplanar dust in our images, though our treatment then also
ignores dust that may be smoothly distributed with the stars.  The
validity of this assumption depends upon the amount of smoothly
distributed dust at various distances from the midplane.  Kylafis \&
Bahcall (1987) estimate the isothermal scale height of dust in NGC 891
to be $z_{dust} \approx 220$ pc, with a visual optical depth through
the center of the galaxy $\tau_o \approx 10.3$.  This implies $\tau_V
(R = 0 \ {\rm pc}; \ z = 500 \ {\rm pc}) \approx 0.4$; though by $z
\approx 700$ pc, this value has decreased to $\tau_V \approx 0.07$.
However, the very structures we are studying in this paper may have
played a role in the derivation of the scale height by Kylafis \&
Bahcall.  If the derived scale height is not appropriate for the
smoothly distributed dust, but is significantly increased by the
presence of the discrete features, any correction we might make to our
analysis would be overly large.  The majority of the features we
discuss in detail are at heights where the adverse effects of ignoring
a smoothly distributed component are likely relatively minor.  At this
point, we will neglect the addition of a smoothly distributed dust
component in our analysis.

We will assume that the dust in NGC 891 has extinction that is
compatible with the parameterization of Cardelli et al.  (1989,
hereafter CCM), which describes Galactic interstellar extinction with
the single parameter $R_V[\equiv A_V / E(B-V)]$.  The Galactic average
for the diffuse interstellar medium is $R_V \approx 3.1$, though this
value seems to vary with environment (CCM, Savage \& Mathis 1979).
Using equation (\ref{eqn:xtau}) and the apparent extinction data in
Table \ref{table:observed} it is possible to constrain the value of
\Rv\ for the features seen in NGC 891.  Figure \ref{fig:extinct} shows
a plot of \ab\ and \ar\ versus \av\ for a sample of 36 dust features,
not all of which will be individually discussed here.  Also plotted
are the predicted values of these quantities for various values of
$x$, using the CCM parameterization with \Rv=3.6, the value of which
defines the slope of the relationships between the various wavebands.
One can see that the data points for \ab\ and \ar\ for \av
$\lesssim$0.7 describe well the trends expected for CCM Galactic
extinction.  Furthermore, this is in a regime where effects of the
uncertain value of $x$ are relatively small.  Therefore it is possible
to constrain the value of \Rv\ with the low-apparent extinction data
for NGC 891.  We have minimized the combined $\chi ^2$ of \ab\ and
\ar\ vs. \av\ about CCM models of differing \Rv.  Using the variance
of the individual measurements of $a_\lambda$ about the mean given for
the features in Table \ref{table:observed}, we estimate the errors in
the apparent extinctions to be $\pm 10$\% ($\sigma_{a_\lambda} \approx
0.1$ mag.).  The minimization of $\chi ^2$ using the estimated errors
in $a_\lambda$ yields \Rv $= 3.6 \pm 0.4$ as the best fit to the data.
This value is roughly consistent with the average \Rv\ for Milky Way
disk clouds.  For illustrative purposes, we have also plotted dashed
lines for \Rv=3.2 and 4.0, $x$=0 on Figure \ref{fig:extinct}, which
represent 1 $\sigma$ deviations from the best fit value of \Rv .

\subsection{Derived Properties of Individual Dust Features}
\label{subsec:derived}

Equation (\ref{eqn:xtau}) may be used to derive information on the
individual features identified in Figures \ref{fig:WIYNcenter} and
\ref{fig:WIYNne}.  Let us assume a CCM extinction law with $R_V
\approx 3.6$, which implies $\tau_B = 1.29 \tau_V = 1.68 \tau_R$.
This allows us to solve for the best values of $x$ and $\tau_V$, using
equation (\ref{eqn:xtau}) in B, V and R.  Table \ref{table:derived}
gives the derived values of the V-band extinction $A_V[= 1.086
\tau_V]$ and $x$ for the individual features discussed in
Sect. \ref{subsec:individual} and identified in Figures
\ref{fig:WIYNcenter} and \ref{fig:WIYNne}.  In general the values of
$A_V$ and $x$ change by $\pm$10-15\% as one changes the assumed value
of \Rv\ by $\pm 1 \sigma$. The greatest contributor to the
uncertainties in these values is likely our neglect of scattering,
however.

\begin{planotable}{cccccc}
\tablenum{3} 
\tablewidth{0pc} 
\tablecolumns{6} 
\tablecaption{Derived Properties of Selected Dust Features} 
\tablehead{ \colhead{Feature} & \colhead{A$_{\rm V}$\tablenotemark{a}}
& \colhead{N$_{\rm H}$\tablenotemark{b}} & $x$\tablenotemark{c} 
& \colhead{Mass\tablenotemark{d}} & \colhead{Energy\tablenotemark{e}}\\
\colhead{} & \colhead{} & \colhead{(10$^{20}$ cm$^{-2}$)} 
& \colhead{} & \colhead{($10^5$ M$_{\odot}$)} 
& \colhead{(10$^{51}$ ergs)} }
\startdata 
1 & 1.79 & 30 & 0.1 & 40 & 130 \nl 
2 & 0.79 & 13 & 0.1 & 2 & 25 \nl 
3 & 0.94 & 16 & 0.5 & 5 & 45 \nl 
4 & 0.71 & 12 & 0 & 8 & 40 \nl 
5 & 0.73 & 12 & 0.3 & 3 & 24 \nl 
6 & 1.00 & 17 & 0.3 & 5 & 20 \nl
7 & 0.66 & 11 & 0.5 & 2 & 30 \nl 
8 & 0.80 & 13 & 0 & 5 & 20 \nl 
9 & 0.90 & 15 & 0.4 & 20 & 80 \nl
10 & 0.85 & 14 & 0.3 & 20 & 70 \nl 
11a & 0.63: & 10: & 0: & 10: & 50: \nl
11b & 1.96: & 32: & 0.7: & 50: & 240: \nl 
12 & 1.26 & 21 & 0.2 & 20 & 120 \nl
\enddata
\tablenotetext{a}{Value derived for the V-band extinction (in
magnitudes) using equation (2) in three colors and assuming
$R_V\approx3.5$.  These estimates neglect the fact that starlight is
scattered into the beam (see Sect. 4.2).}
\tablenotetext{b}{Approximate column density of material (in units of
$10^{20}$ cm$^{-2}$) assuming Galactic extinction and gas-to-dust
conversions.}  
\tablenotetext{c}{Fraction of stellar light coming from in
front of the absorbing feature derived using equation (2) in three
colors. This quantity is to be considered highly uncertain,
particularly given our neglect of scattered light (see Sect. 4.2).}  
\tablenotetext{d}{Approximate mass based upon column density and
projected area (given in units of $10^5$ M$_\odot$).}  
\tablenotetext{e}{Approximate potential energy of the observed
material given its mass and height from the midplane (in
units of $10^{51}$ ergs, the approximate energy imparted to the ISM
from a single supernova).}
\end{planotable}

Bohlin et al. (1978) used the {\em Copernicus} observatory to show
$N(\mbox{\ion{H}{1}}) + 2N({\rm H_2}) \equiv N_H \approx 5.8 \times
10^{21} \ E(B-V)$ [cm$^{-2}$] in the disk of the Milky Way.  Here
$N(\mbox{\ion{H}{1}})$ and $N({\rm H_2})$ are the \HI\ and H$_2$
column densities, respectively, and $N_H$ is the total hydrogen column
density, not including the contribution from ionized gas.  To estimate
the amount of gas associated with the absorbing dust structures, we
apply this relationship between gas and dust to the extraplanar
features in NGC 891.  Assuming \Rv$\approx$3.6, we use
\begin{equation}
N_H = 1.7 \times 10^{21} \ A_V \ 
	{\rm [cm^{-2}]}
\label{eqn:columndensity}
\end{equation}
to estimate the column density of associated gas in the observed dust
structures.  This estimate, when coupled with the approximate
projected area of a given structure, yields a mass estimate of each
feature; both quantities are given in Table \ref{table:derived} for
the features discussed herein.  The mass estimate given here includes
a correction factor of 1.37 to convert from the mass of hydrogen to
the total mass accounting for typical fractions of helium and the
heavy elements.  The values of \Nh\ derived using equation
\ref{eqn:columndensity} are only appropriate if the gas-to-dust ratio
in the observed structures is approximately that of Milky Way disk
clouds; this will be an underestimate of the column density if
significant dust destruction has occured, an overestimate if the dust
has been separated from the gas (see Sect. \ref{sec:discussion}).  Our
estimates for the column density of material vary by $\pm$15-20\% as
the value of \Rv\ is changed by $\pm 1 \sigma$.

Except in cases of significant \Ha\ contamination in the R-band
images, the measurements we have made will tend to underestimate the
apparent extinctions due to difficulties with the definition of the
background surface brightness, $S_{bg, \lambda}$, and thus
underestimate the optical depth and column density.  The quantities
given in Table \ref{table:derived} for features whose extinction
measurements may be contaminated by \Ha\ emission are marked with a
colon, indicating a greater degree of uncertainty in the derived
values of $A_V$, \Nh\ and mass.  We further underestimate the optical
depth due to the effects of light scattered into the line of sight by
the dust.  Our estimates of the quantities given in Table
\ref{table:derived} are therefore to be considered lower limits.

In general our estimates of the gas column densities associated with
the observed dust absorption give $N_H \gtrsim 10^{21}$ cm$^{-2}$.
These estimates place the individual features in the high end of the
realm occupied by diffuse clouds in the disk of the Milky Way.  If we
assume the depths of the features are similar to their widths, we
obtain density estimates of $\langle n \rangle \sim 3-20$ cm$^{-3}$.
Many of the features are likely to include molecular material if these
estimates of column and particle densities are correct.  For Milky Way
disk gas, the average value of the molecular fraction [$f \equiv 2
N({\rm H_2}) / N_H]$ for material towards stars having $0.3 < E(B-V) <
0.53$, or $0.9 \lesssim A_V \lesssim 1.6$, appropriate for most of the
structures in Table \ref{table:derived}, is $\langle f \rangle \approx
0.22$ (Savage et al. 1977).  Thus nearly one-quarter of the hydrogen
content of the extraplanar features may be tied up in molecules, if
the gas content and physical conditions of the dust structures are
similar to those in the disk of the Milky Way, a possibly dubious
assumption.  The predicted masses for these features are quite large:
the masses given in Table \ref{table:derived} are similar to those for
the Galactic giant molecular clouds.  The sizes of most of the
structures discussed here, however, are significantly larger than
those for typical GMCs (sizes $< 100$ pc).

If the hydrogen in these structures is mostly atomic, the values of
\Nh\ given in Table \ref{table:derived} are above the detection limits
of the \HI\ surveys of Rupen (1991) and Sancisi \& Allen (1979);
however, the individual features discussed here would be unresolved
and almost impossible to identify.  Indeed, some of the features
apparent in our WIYN observations are washed out in images with seeing
worse than 1.0\arcsec.  The CO observations of Scoville et al. (1993)
and Handa et al. (1992) begin to be able to resolve some of the
structures.  A comparison of our images with the total intensity map
of Scoville et al. shows some correspondences near the bases of
several dust structures (e.g., feature 1), though the CO data do not
show material much above $|z| \sim 500$ pc.  The correspondences may
simply be coincidence.  Unfortunately, the molecular spur discussed by
Handa et al. (1992) lies near the bright star seen just above the
plane in Figure 2(b); we are not able to measure reliable surface
brightnesses for that feature given the slope of the background light
and the contamination from the bright foreground star.
Garc\'{\i}a-Burillo et al. (1992) obtained several observations
towards directions that showed dust features extending beyond the thin
disk of NGC 891; they find no clear difference in the CO properties of
sightlines on or off of these dust structures, though they sample few
of the prominent dust features.  Their spectra towards dust features
have broad velocity widths and suggest a CO component extended along
the line of sight, which they argue may be evidence for CO emission
independent of the vertical dust features. The tangle of structures
seen between the prominent features discussed here is evidence for
dust filaments far along the line of sight into the low-halo of the
galaxy.  The CO observations may therefore be detecting many
structures along the path through the thick disk.

Given the derived extinctions and column densities given here, which
imply quite thick dust structures, we might expect to be able to
observe thermal emission from the dust itself.  Gu\'{e}lin et
al. (1993) have mapped NGC 891 at 1.3 mm using the IRAM 30-m
telescope.  Their data show cold dust along the major axis of the
galaxy, but they do not mention the vertical extent of the material,
nor does there appear to be significant thermal emission in their map
associated with the most obvious dust features seen in our images.
NGC 891 was observed with IRAS using the Chopped Photometric Channel.
The emission from NGC 891 at 50 and 100 $\mu$m was unresolved
perpendicular to the plane with this instrument, which at these
wavelengths has a resolution of $\sim$75\arcsec\ and 85\arcsec\,
respectively (Wainscoat et al. 1987).  In general observations that
may have shown thermal emission from extraplanar dust have been
hampered by insufficient resolution or sensitivity to measure dust
beyond the thin disk of the galaxy.

Using the masses given in Table \ref{table:derived}, we may make
estimates for the gravitational potential energies of the observed
structures given their positions above the midplane.  Approximating
the distribution of mass density using the isothermal sheet model
($\rho \propto {\rm sech}^2 (z)$; though see van der Kruit 1988), the
potential energy $\Omega$ of a structure at a height from the midplane
$z$ is
\begin{equation}
\Omega=10^{52} {\rm ergs} \left( \frac{M_{cloud}}{10^5 {\rm M_\odot}}
	\right) \left( \frac{z_o}{700 {\rm pc}} \right) ^2 \left(
	\frac{\rho_o}{0.185 {\rm M_\odot pc^{-3}}} \right) \ln [ \cosh (
	z/z_o ) ]
\label{eqn:energy}
\end{equation}
where $M_{cloud}$ is the mass of the dust cloud, $z_o$ is the mass
scale height of the (stellar) disk and $\rho_o$ is the mass density at
the midplane.  The value $z_o = 700$ pc used here is from a fit to the
stellar light in near-infrared images of NGC 891 (Aoki et al. 1991),
which we will assume approximates the mass density scale height, and
we have adopted the mass density $\rho_o = 0.185$ M$_\odot$ pc$^{-3}$
appropriate for the solar neighborhood (Bahcall 1984).  The
relationship given in equation \ref{eqn:energy} is likely not
appropriate for material very near the midplane or beyond $z \approx
1.5$ kpc, with the high end being more uncertain (Aoki et al. 1991;
Kuijken \& Gilmore 1989).  Though this series of approximations may
not lead to highly accurate potential energies, it is interesting to
examine these order of magnitude estimates.  It should be noted that
this equation gives only the potential of the material at a given
height $z$.  It makes no assumptions as to the fraction of input
energy that is radiated or converted to purely thermal motion, which
would certainly be important in most expulsion scenarios (see
discussion in Sect. \ref{sec:discussion}).  Thus if used to
approximate the input energy required to lift the material from $z=0$
to its observed height, the energies derived must be treated as lower
limits.  Further, no kinematic information for these structures
exists, so no estimate of the kinetic energies of these dusty
structures can be made, again making the energies calculated with
equation \ref{eqn:energy} lower limits to the total input energy.
Indeed, these structures may be at mid-points in their evolution, and
the processes responsible for forming these structures may still be
depositing energy into them, adding even more energy to the mix.  In
Table \ref{table:derived} we give the approximate energies derived
using equation \ref{eqn:energy}.  For structures extended in $z$ we
have used a central value for the $z$-height of the feature in
deriving the energy. The energies given in Table \ref{table:derived}
are quite large, but similar to those derived for supershells in the
Milky Way (Heiles 1979).  Those values given in Table
\ref{table:derived} are equivalent to the kinetic energy input into
the ISM of tens to hundreds of Type II supernovae, though the input
energy required to lift the observed features from the midplane is
greater than this due to radiative and thermal losses.

\subsection{Notes on Individual Features}
\label{subsec:individual}

We have derived the properties for a set of individual dust features
seen far from the plane of NGC 891.  We have not attempted an unbiased
selection of the individual structures, but rather have chosen some of
the more ``interesting'' and clearly visible features.  In general we
have selected relatively small structures that are more likely to
represent coherent structures and are relatively easily seen in
hardcopies of the images.

Our analysis of these individual dust features gives striking numbers.
The features are characterized by by $N_H \gtrsim 10^{21}$ cm$^{-2}$,
with masses estimated to be more than $2 \times 10^5 - 5 \times 10^6$
M$_\odot$, assuming Galactic gas-to-dust relationships.  These numbers
are quite large, and may imply the presence of a significant amount of
molecular material, as mentioned earlier.  The gravitational potential
energies of the individual dust structures are similarly large:
$20-240 \times 10^{51}$ ergs, which represents the energy input of at
least tens to hundreds of supernovae, possibly an order of magnitude
more.

The absorbing structures observable far from the plane of NGC 891 have
a wide range of morphologies, and even of physical properties.  One of
the more extreme examples of these individual features is feature 1,
which lies to the SW of the galaxy's center.  The derived mass
estimate places it at the high end of the mass range for the Galactic
giant molecular clouds.  It is quite opaque and may harbor a
significant amount of molecular mass.  There is some evidence for a
slight $z$-extension of CO gas at the base of this feature in the maps
of Scoville et al. (1993).  Very deep interferometric CO observations
of this galaxy might yield emission from high-$z$ dust features such
as this one.

Feature 7 is quite interesting for its morphology; it appears as a
cometary-shaped absorbing feature 1.4 kpc from the mid-plane.  There
is also lightly-absorbing material seen ``trailing'' this feature at
higher $z$.  There are several other features at similar $z$-heights
that seem to have similar morphologies; unfortunately, the apparent
extinctions in the V and R bands for many of these are too low to
measure in the current data set.  Deeper observations will allow us to
probe dust higher into the halo of NGC 891.

Among the most interesting of the individual structures, particularly
in the context of the disk-halo interface in galaxies, are features 10
and 11.  These structures lie in the NE portion of the disk of NGC 891
and are almost certainly connected with the star-formation activity on
this side of the galaxy.  This region shows enhanced \Ha\ emission, as
well as several structures whose shapes are reminiscent of supershells
(Rand et al. 1990; Pildis et al. 1994a).  Given the known presence of
\Ha\ emission near these two features, though seemingly slightly
offset, the quantities derived for these two structures should be
considered more uncertain than those derived for the previously
discussed features due to \Ha\ contamination in the R-band image.

Feature 11, which appears as an arc open to high $z$, outlines
an \Ha\ emitting ionized shell quite well.  The ``walls'' of the dusty
shell are approximately 50-60 pc thick with a maximum separation of
almost 600 pc.  The center of the shell is itself $\sim$ 600-650 pc
above the midplane.  The two sides of this structure have different
absorbing properties.  In Table \ref{table:observed} we refer to the
southern-most side, which is closest to the bulge of the galaxy, as
11a and the northern-most as 11b.

Feature 10 appears as a cone of dusty material, reaching to a maximum
traceable height of $z \approx 900$ pc.  The apparent opening angle of
this feature is $14^\circ$.  Ionized gas appears brighter along the
exterior of the cone-like structure, tracing its outer boundary very
well.

To show the relationship between the ionized gas structures and the
absorbing dust features, we compare the B-band image and an \Ha+\NII\
image of the NE section of NGC 891 in Figures \ref{fig:halpha}(a) and
(b).  These relatively shallow 0.9\arcsec\ resolution images were
taken at WIYN through a filter not optimized for the redshift of NGC
891 ($\lambda_{cen} = 6570$ \AA, FWHM = 16 \AA).  The response to \Ha\
changes as a function of position along the disk, given the rotation
of the galaxy, and the relative throughput of \Ha\ vs. \NII\ also
varies along the disk.  Even given the non-optimal filter
characteristics, the image is useful for comparison with the dust
structures seen in this portion of the disk.  One can see structures
in ionized gas associated with the dust features 10 and 11,
particularly the almost circular bubble-like feature that is outlined
by the dust of feature 11.  The edges of the cone-like feature 10 are
also bright in ionized gas emission.  This region of the galaxy shows
very pervasive ionized gas.  The star-formation rate is thought to be
very high in this section of the disk, and our data reveal a plethora
of ionized regions that are powered by young stars in the disk.  The
color image shown in Figure \ref{fig:WIYNcolor} shows that these
regions appear quite blue, also indicating the presence of massive
stars.

The \Ha+\NII\ images reveal that some of the dust structures observed
in NGC 891 do have counterparts tracable in ionized gas, while others
show no detected ionized gas emission.  There is a quite robust
association of features 10 and 11 with energetic processes driven by
the existence of hot stars.  An in-depth comparison of the
relationships of these and other structures with counterparts seen in
ionized gas will require deeper \Ha\ images through a more appropriate
filter and will be the subject of future work.

\subsection{The Dusty Thick Disk}
\label{subsec:thickdisk}

While we have concentrated our discussion thus far on some of the more
prominent features, a large ensemble of less distinct or smaller
high-$z$ extinction structures are apparent in the WIYN images.
Number counts of these features are difficult to make because we have
no objective criteria for what constitutes an individual structure;
chance superposition of several clouds along the line of sight through
the galaxy are almost certainly producing some of the structures we
might by eye identify as singular features.  We are also unlikely to
pick up low column density material or features that lie very far into
the galaxy; indeed, we would be unable to identify even large optical
depth features with $x \gtrsim 0.6 - 0.7$.  We have, however,
attempted a rough number count of high-$z$ dust features in the halo
of NGC 891.  We make no claims as to the completeness of such a count.
We have been relatively conservative in identifying structures, and
therefore our counts represent a lower limit to the number of
features.  Our rough number count for structures having $\vert z \vert
\gtrsim 400$ pc and $a_B \gtrsim 0.25 - 0.3$ mag. yields more than
than 120 features.  If these features exist on the front side of the
galaxy, with no intervening stars, then $a_B \sim A_B$ and $N_H
\gtrsim 4 \times 10^{20}$ cm$^{-2}$ for these features, neglecting
scattering.

Counts taken along lines at $z = \pm 500$ yield approximately 30
features ($a_B \gtrsim 0.3$) on each side of the galaxy within the
central $\pm 3$ kpc projected distance from the center of the galaxy;
similar results are obtained for counts at $z = \pm 700$ pc.  There is
no evidence for an asymmetry in the number of counts across the
midplane of the galaxy, as one might expect if a warp were the cause
of these dust features.  Unfortunately, comparisons between the number
of features per kpc in either the radial or vertical directions are
severely hampered by the exponential decay of the background
illumination.

The 12 structures identified in Figures \ref{fig:WIYNcenter} and
\ref{fig:WIYNne} together contain $\sim 2 \times 10^7$ M$_\odot$ of
material, if the assumption of Galactic gas-to-dust extinction ratios
is appropriate.  If the 120 features we count have properties similar
to those discussed above, then the mass tied up in the visible
structures with $|z| \gtrsim 400$ pc may be $> 2 \times 10^8$
M$_\odot$.  This represents $>$2\% of the total mass of neutral
gas, \HI+H$_2$, in NGC 891, estimated to be $\sim 9 \times 10^9$
M$_\odot$ (Garc\'{\i}a-Burillo et al. 1992).

It is clear that there are a large number of absorbing structures far
from the mid-plane of NGC 891 that may contain a significant amount of
mass.  We associate this material with the known thick disks of
neutral and ionized hydrogen in NGC 891 (Swaters et al. 1997; Dahlem
et al. 1994; Dettmar 1990; Rand et al. 1990).  This association is
only based upon the fact that the dust far from the plane shows a
similar extent to the gaseous components of the thick interstellar
disk of NGC 891.  Our images show that the dust associated with the
thick disk is quite structured, not smoothly distributed.  Indeed, the
\Ha\ thick disk is broken up into filamentary structures (Dettmar
1990; Rand et al. 1990).  However, deep high angular resolution
observations of \Ha\ emission do not exist.  Therefore, a detailed
comparison between the extraplanar dust structures and \Ha\ structures
will require new observations.

In future work on this galaxy, we will attempt to better our
quantitative understanding of the distribution of dust, particularly
searching for evidence of a two-component distribution of the dust.
The existence and properties of a dusty thick disk are important for
understanding the processes that eject material far from the plane and
the relationship of the dust to other tracers of the interstellar
thick disk, as well as for correcting for the effects of dust upon
other observations (e.g., derivation of \Ha\ scale-heights).  However,
the derivation of the thick disk dust scale height is rife with
difficulties, and hence we have not attempted it in this first paper
on the extraplanar dust of NGC 891.  The derivation of such a scale
height is heavily dependent upon a correct model for the distribution
of stellar light in the galaxy, particularly for the presence of a
stellar thick disk (see Morrison et al. 1997).  Correct fitting the
observed light distribution to determine the distribution of dust also
requires a good treatment of the radiation transfer through the
galaxy.  Kylafis \& Bahcall (1987) have studied the distribution of
stars and dust in NGC 891, though they assume a one-component model
for the dust.  Whether adding a thick dusty disk better fits the
photometric data remains to be seen.

\section{HST Images and Seeing Effects}
\label{sec:hst}

NGC 891 was observed by HST in the imaging snap-shot survey of
Illingworth et al. (HST Prop. 5446).  We have retrieved the Wide
Field/Planetary Camera 2 (WFPC2) images from the HST archive; the
images cover the central region of the galaxy and are useful for
comparing with our WIYN images.  Table \ref{table:log} gives the
properties of the HST WFPC2 images.  The two 90-s exposures taken with
the F606W filter were co-added and mosaiced within IRAF; the final
WFPC2 image is presented in Figure \ref{fig:wfpc2}.  The resolution of
the image is 0.1\arcsec, and is therefore a good measure of the degree
of degradation caused by seeing effects in the WIYN images.

We have measured the widths (short axis) of all of the features
identified in Table \ref{table:observed} that are contained within the
WFPC2 image (features 1, 2, 3 \& 4) as well as several other features
common to the images.  We use the WIYN V-band image for these
measurements, which is a good match to the F606W filter used in the
HST observations (see below).  The measurements of features in the HST
images agree with those measured in the WIYN images to about one or
two WFPC2 pixels, or 5-10 pc.  This correspondence is excellent
considering the general complexity of the absorbing features.

While the HST image shows the features on a finer scale, the bulk of
the apparent structure in the most prominent absorbing regions is
resolved by WIYN.  Some of the features show more filamentary edges
and overall structure in the HST image.  There are a few instances of
features that are washed out by the seeing at WIYN but are detected in
the HST image.  However, there are also low-contrast features at high
$z$ that are detected in the WIYN images but not the HST image.  The
HST image more clearly resolves into individual features some of the
absorbing material that lies further along the line of sight into the
galaxy.  As discussed above, there still exists a tangle of
overlapping features that can be picked out, but not distinguished, in
the HST image.

To more clearly show the difference between the WIYN and HST views of
NGC 891, we present close-up views of the WIYN B-band and HST images
for features 1 and 2 in Figure \ref{fig:feat1} and for features 3 and
4 in Figure \ref{fig:feat4}.  The HST images do show that there are
structures on finer scales than can be detected with WIYN; however,
the bulk of the structure seems to be apparent in the WIYN images.
Though the HST image provides a factor of six better resolution, the
larger features identified in the WIYN images do not break down into
separate structures in the HST image. While they may still be
superpositions of several clouds along the line of sight, few if any
of the structures that seem to be connected in the WIYN images are
separable into distinct structures at the resolution of the HST image.
Feature 4 is an example of this (see Figure \ref{fig:feat4}).  In the
WIYN image, it gives the appearance of being a mostly-coherent
structure stretching over several hundred pc.  The HST image reveals
nothing that contradicts this immediate impression.

The WFPC2 F606W filter is centered at $\lambda=5957$ \AA,
approximately midway between the centers of the V and R filters used
with WIYN, but is very broad, having approximately the same
short-wavelength cutoff of the V filter.  A convolution of the F606W
filter transmission curve with the expected flux distribution produced
by $T_{eff}$ = 5,000 and 10,000 K stars subjected to CCM extinction
curves yields $\langle A_{F606W} / A_V \rangle \approx 0.97$,
independent of $R_V$ (using Kurucz 1991 atmospheres; see Cole et
al. 1991).  This is not particularly surprising since, although the
F606W filter extends much further into the red than does the V-band
filter, the extinction over this bandpass is dominated by the
short-wavelength end (at least for a flat-spectrum source).  We have
measured the apparent extinctions of many of the features common to
both the WIYN V-band and HST F606W images as prescribed in the last
subsection, though we have used 10 pixel wide cuts for the HST image
to account for the difference in pixel size.

As discussed above, the result is a spatial average of the apparent
extinction for a given feature.  Using the standard deviation of the
individual cuts about the mean for a feature as an estimate of the
error, our measurements of absorbing properties for structures in both
images typically agree within one standard deviation.  There are,
however, a few features for which this is not the case.  These are
typically characterized by having a ``core/halo'' structure, i.e., a
central very opaque feature surrounded by a less-absorbing region.
The discrepancy in apparent extinction arises because we adopt the
minimum brightness ratio, $S_{dc,\lambda} / S_{bg, \lambda}$, as
representative of the whole.  When deriving the brightnesses from the
WIYN images, our individual measurements have a similar resolution in
both dimensions, using cuts with widths of 5 pixels combined with a
seeing-limited resolution of 3-4 pixels in the perpendicular
dimension.  The HST measurements are less likely to be an average in
both directions, having cuts of 0.1\arcsec$\times$1.0\arcsec\ through
each feature.  Measurements of features in the HST image, given our
technique, sample the most opaque regions of these ``core/halo''
structures and are less an average of the cloud as a whole.  Though
our measurements of apparent extinctions for various features show
some discrepancies, we believe the method generally gives information
appropriate for deriving the spatially-averaged properties of the
absorbing structures.

\section{Discussion}
\label{sec:discussion}

From the images presented herein it is immediately clear that clouds
of dust, if not dust and gas, exist at large distances from the
midplane of the spiral galaxy NGC 891.  It is not known how common
these extraplanar dust clouds are in spiral galaxies, nor is it clear
that the absorbing structures seen in this galaxy are all caused by
the same phenomenon; even given these uncertainties, it is important
to begin a discussion of mechanisms that may be responsible for the
features we see in the WIYN and HST images of NGC 891.  Although small
quantities of dust at large distances from the planes of galaxies can
be produced by direct injection from evolved stars and supernovae, it
appears for galaxies like NGC 891, with extensive dust structures at
large $z$, that some type of ejection process from the reservoir of
material in the disk is required to produce the amount of high-$z$
material seen.  Further, the complex morphologies of the structures
seen in our data strongly suggest mechanisms other than direct
injection for producing these absorbing features.  In the following we
discuss four types of ejection processes: 1. Hydrodynamical phenomena,
including fountain flows. 2. Expulsion by radiation pressure on dust.
3. The effects of magnetic fields, and 4. ejection mechanisms
involving dynamical instabilities.

\subsection{Ejection via Hydrodynamical Phenomena}

It may be possible to transport dust from the thin disks of spiral
galaxies via hydrodynamical phenomena such as ``galactic fountain''
flows (Shapiro \& Field 1976; Bregman 1980; Houck \& Bregman 1990) or
``galactic bores'' (Martos et al. 1996, cf. Suchkov \& Shchekinov
1975).  The expulsion of dust in these scenarios would seem to depend
sensitively upon the physical conditions at the sights of expulsion,
e.g., supershells for the fountain model.  In the fountain model there
are two possibilities for lifting dust beyond the thin disks of
spirals.  The first relies upon the upwelling hot gas to carry the
dust.  This mechanism seems unlikely to be able to transport a large
amount of dust, either by entrainment of material as the hot gas
escapes the disk or by feeding the dust into the hot interiors of
supershells prior to the expulsion of the hot material, due to the
extremely low density of the hot gas and the harsh conditions within
this hot medium.  Ferrara et al. (1991), in their numerical models of
dust behavior when immersed in a hot coronal halo, find that the dust
is able to move relatively unimpeded in such conditions.  The inverse
is likely also true: when the hot gas is expanding out of the disk of
the galaxy, it may not offer much lift to the dust.

Possibly more likely is the transport of dusty material from the thin
disk within the dense walls of supershells.  While the initial
expansion of a supernova remnant or superbubble (SNR) may cause strong
shocks, through most of their evolution SNRs expand at relatively low
velocities.  In this slowly moving phase a remnant sweeps up
considerable mass which does not experience shocks with large velocity
jumps.  Thus the dust swept up in this phase of SNR evolution may
remain relatively intact, i.e., the shocks dust may encounter in this
phase are not sufficiently strong to destroy large fractions of the
grains.  As a superbubble expands into the halo, it transports this
undestroyed dust upwards within its walls.  The chimney model of
Norman \& Ikeuchi (1989) suggests that if the supershells are able to
break out they will leave remnant, kiloparsec-scale
vertically-oriented walls, which may be dust-laden.  Indeed, Koo et
al. (1992) have identified many candidate Galactic ``worms'' in both
\HI\ and IRAS 100 $\mu$m emission that may be such remnants or the
walls of supershells themselves.  The absorbing features seen in our
images may be pronounced examples of these dusty worms, particularly
the features found in the NE section of the disk.  Feature 11, for
example, seems to be the dusty wall of a known \Ha\ supershell.

Dense knots of material which follow ballistic trajectories may also
be expelled from the disk by superbubbles.  Knots formed by
Rayleigh-Taylor instabilities as a superbubble breaks through the disk
material and begins to accelerate into the halo follow ballistic
trajectories away from the disk, as they have significantly higher
column densities than the halo material (Cioffi 1986).  These knots of
material may lead to relatively disconnected clouds of dense material
in the low-halo of spiral galaxies.  If dust survives in these
fragments of the evolved superbubble, the ejected clouds may provide a
source of halo dust grains.  Greater knowledge of the degree to which
dust can survive in the walls of supershells is needed to assess the
role these structures may play in lifting dust beyond the thin disk of
spirals.

Models which predict upwelling gas trajectories as material ``falls''
into the enhanced gravitational potential of a spiral arm, such as the
``galactic bore'' model of Martos et al. (1996), may also exhibit a
sensitive dependence upon the prevailing local conditions.  The depth
of the gravitational potential well into which the material falls and
the velocity with which gas encounters the spiral density wave would
both be a determining factor in the resulting temperature of the gas
(Kovalenko \& Levy 1992).  Thus the prevailing physical conditions
where the upwelling gas is expected to be produced would depend upon
the location of the gas with respect to the corotation radius, since
the velocity with which gas enters the density wave depends upon its
location relative to corotation, and the amplitude of the density
wave.  Dust grains in material that encounters too strong a shock as
it falls into the enhanced gravitational potential of a spiral arm may
be too efficiently destroyed to produce substantial high-$z$ opacity,
even if the gas-dust drag were sufficient to lift the grains as the
heated gas expands upwards into the low halo.

The literature on the effects of either model (fountain vs. bore) on
dust grains and their distribution is quite sparse.  Much work needs
to be done, both observationally and theoretically, before a
reasonable discrimination can be made between these types of models
and those that rely on other processes to lift grains from galactic
thin disks.

\subsection{Ejection via Radiation Pressure}
\label{subsec:radiative}

Ferrara et al. (1991), Franco et al. (1991) and Ferrara (1993) have
suggested that radiation pressure on dust grains can play a
significant role in the kinematics of both the gas and dust in spiral
galaxies.  Franco et al. (1991) investigated the effects of radiation
pressure on dusty clouds in the thick interstellar disk of the Milky
Way, and by extension other spirals.  Their results suggest that
interstellar clouds having gas-to-dust ratios typical for Milky Way
clouds can be ``photolevitated'' to a few hundred parsecs above the
midplane of the Galaxy.  Clouds directly above luminous star clusters
or near spiral arms may be lifted to even greater heights.  Ferrara
(1993), extending this work, suggested the vertical support of the
neutral interstellar thick disk of the Milky Way (Lockman 1984) could
be provided by turbulence when including the stellar radiation field,
which lowers effective gravitational potential of the dusty material.

In contrast to the studies above, which discuss the distribution of
clouds near the midplane, the calculations of Ferrara et al. (1991)
follow the trajectories of dust grains immersed in a hot hydrostatic
galactic corona.  They find that dust can be expelled from a model
Milky Way disk to quite large distances ($z \sim 20-120$ kpc).  It
should be noted, however, that they include no diffuse clouds beyond
the $z=200$ pc initial position of the grains. Diffuse clouds above
the plane, even if they fill a small fraction of the grains'
trajectories through the halo, could cause significantly more drag
than the models of Ferrara et al.  Warm neutral clouds are known to
exist at large $z$-heights in our Galaxy.  The thick disk of neutral
material or ``Lockman layer'' discussed above extends to $z$-heights
in excess of 500 pc (Dickey \& Lockman 1990), while intermediate
velocity clouds may lie significantly higher (e.g., Wakker et
al. 1996).  However, the influence of magnetic fields may be the most
significant modifier to the dust trajectories of the Ferrara et
al. models (see below).

More recently Ferrara \& Shull (as discussed in Ferrara 1997) have
begun modelling the time evolution of the dust distribution in the
immediate vicinity of OB associations via Monte Carlo models.  Their
models account for the more extended $z$-component of \HI\ and for the
evolution of the radiation field of the OB association.  They find
that dust can be expelled well beyond the thin interstellar disk of
their model galaxy and predict a relatively flat distribution of dust
grains above these collections of stars.  The models further predict
that small grains tend to be expelled to greater distances than large
grains.  In the early evolution of their ``dusty chimneys,'' the
distribution of dust in their simulations is roughly conical, with the
base of the cone pointing away from the midplane; this morphology is
not unlike that of feature 10.  This work seems quite promising, for
it includes the influence of the interstellar thick disk and makes
predictions that should be testable with high spatial resolution
infrared images of edge-on galaxies (e.g., with SIRTF or SOFIA, which
could be used to search for evidence for a varying degree of dust
destruction or changing gas-to-dust ratios with $z$).

A concern with models that rely upon radiation pressure to expel
material from the disk of a galaxy is the relatively large optical
depths we have derived for the high-$z$ features identified in Figures
\ref{fig:WIYNcenter} and \ref{fig:WIYNne}.  Dust clouds that are
self-shielding would seem to be difficult to drive radiatively and may
be subject to radiation-induced instabilities.  Indeed, Franco et
al. (1991) argue that clouds having $N_H \gtrsim 5 \times 10^{20}$
cm$^{-2}$ or $A_V \gtrsim 0.3$, assuming Milky Way gas-to-dust ratios,
are unable to be driven effectively.  This is at least a factor of two
below our lowest column density/extinction estimates given in Table
\ref{table:derived}.  Those features that stretch as ``columns''
vertically away from the plane may have even larger column densities
and optical depths when viewed from the disk of NGC 891 (e.g.,
features 1 and 4).

In general radiative expulsion models predict lesser amounts of grain
destruction than those methods that rely upon violent means, such as
the fountain-type models, though little work has been done
theoretically to describe the extent to which dust is transported or
destroyed in supershells.  Ferrara et al. (1991) have calculated that
bare silicate grains lose some 20\% of their mass in their $\sim2
\times 10^8$ yr trip through a model galactic corona, with graphite
grains losing even less; this time scale is similar to that of total
destruction for grains in the disk of a galaxy which are exposed to
the passage of multiple supernova shocks (Jones et al. 1994).  Jones
et al. (1994) calculate the fraction of grains destroyed in the
passage through a strong shock is $\leq$0.29 and $\leq$0.45 for
graphite and silicate grains, respectively.  If grains pass through a
strong shock and find themselves in the harsh hot interior of a
supershell, the fraction of grain destruction may be greater than this
value.

A further difference between the Ferrara et al. (1991) predictions for
the destruction of grains in a hot halo, and for the destruction one
might envision in a fountain-like scenario, is that the destruction of
grains in the radiatively driven scenario primarily occurs far from
the disk of the galaxy.  In a supernova-driven expulsion scenario, the
bulk of the dust destruction should occur within the first kpc.  A
clear examination of the distribution of elemental depletions with
distance above the plane of our own Galaxy may help to distinguish
between these two types of models.  Abundance studies with HST
(Sembach \& Savage 1996) reveal that grain cores survive the processes
that lift matter to $z$ distances ranging from 0.5 to 1.5 kpc in the
halo of the Milky Way.  Indeed, our WIYN images of NGC 891 provide the
same information for an external galaxy: grains survive the trip from
the galactic plane to the low-halo, by whatever means they are
expelled.  These observations imply that the cores of dust grains are
very resilient and survive while being pushed into the halo by violent
processes, or reach high-$z$ by less extreme means, such as via
radiation pressure.

Knowledge of the dust properties of supershells in the Milky Way and
in other galaxies may constrain the amount of dust that is destroyed
in these structures and help determine whether they are able to expel
dust from the thin disks of spirals.  It may be possible to search for
evidence of grain destruction in the dust features we have discussed
here using emission line imaging or spectroscopy of the near-infrared
[\ion{Fe}{2}] and Pa$\beta$ lines.  The [\ion{Fe}{2}]/Pa$\beta$ ratio
is known to be enhanced in regions of supernova activity as the dust
is destroyed, freeing Fe from the solid phase (Greenhouse et
al. 1991).  The presence of enhanced [\ion{Fe}{2}] emission (over
typical \ion{H}{2} regions, say) would be an indicator of dust
destruction and likely supernova activity.  High-resolution images
would be required to search for emission from specific features; HST,
with the newly installed NICMOS instrument, might be useful for doing
exactly this.

\subsection{The Effects of Magnetic Fields}

Recent radio continuum observations at 2.8 cm (Dumke et al. 1995; see
also Sukumar \& Allen 1991), where the effects of Faraday
depolarization are likely very minor, suggest that the magnetic field
geometry in NGC 891 is predominantly parallel to the plane, on the
large scale.  The orientation of visible and near-infrared
polarization vectors in NGC 891 suggest, however, a vertical magnetic
field if the polarization is due to aligned grains (Scarrott \& Draper
1996; Wood \& Jones 1997).  Again, this discrepancy may be due to the
differing resolutions of the two datasets or to optical depth effects
in the optical/IR region.

The presence of magnetic fields, particularly parallel to the plane,
may significantly affect the processes that cycle material between the
disks and halos of spiral galaxies, though we have thus far neglected
the role of magnetic fields in our discussion.  The addition of
magnetic pressure in the calculated evolution of SNRs generally leads
to smaller remnants, at least in the direction perpendicular to the
local magnetic fields (cf., Slavin \& Cox 1992).  Magnetic fields
oriented parallel to the plane of a spiral galaxy provide a tension
that could confine the $z$-expansion of superbubbles.  This tension
makes it more difficult for large superbubbles to break out of the
thin interstellar disk of a galaxy; all of which would seem to suggest
that the presence of magnetic fields might quench disk-halo
circulation in disk galaxies if of sufficient strength and the correct
orientation.

On the other hand, a configuration wherein horizontal magnetic fields
provide partial support against a vertical gravitational field is
subject to the Parker instability (Parker 1966).  As a supershell or
other disturbance pushes the field upward in a small region, matter
slides down the field lines, which reduces the weight providing
confinement of the magnetic field.  The field continues its expansion
to higher $z$, given that less material is holding it down, forming an
$\Omega$-shaped structure as matter continues to slide down the
now-vertical sides of the magnetic arch.  This process, known as the
Parker instability, may play an important role in the dynamics of
interstellar material (Basu et al. 1997; Matsumoto et al. 1990).
Though magnetic fields are generally thought to impede the blow-out of
superbubbles into the halo, the existence of a Parker-unstable region
of a spiral may in fact aid in the lifting of material beyond the thin
disk (Kamaya et al. 1996).  The Parker instability may be excited
variously by supernovae-driven phenomena (Kamaya et al. 1996) or by
inflation of relativistically hot cosmic-ray gas (Parker 1992).

Indeed, a subset of the absorbing structures seen in NGC 891 is
reminiscent of magnetically shaped phenomena.  Further, the Parker
instability predicts that matter will flow down the sides of the
arched magnetic field.  The material that has fallen down either side
of the field structure will appear as a vertically-oriented
``column.''  Structures such as feature 4 could represent the end
result of this type of magnetic field evolution.

The orientation of the magnetic field could also significantly affect
the expulsion of dusty material via radiation pressure from galactic
starlight.  If the magnetic fields are predominantly parallel to the
plane of a galaxy, it would seem difficult to radiatively expel grains
from the thin disk, assuming the grains are charged and would thus be
coupled to the magnetic field.  In regions where the magnetic field
runs perpendicular to the disk, this would be less significant a
problem.  Sofue et al. (1994) have suggested that the absorbing
structures seen in NGC 253 may be caused by radiation driving along
vertical magnetic fields.  They argue that magnetic fields should be
invoked because of the general morphologies of their features,
particularly given the coherency exhibited by their features over
lengths of a kpc or more.

Though the radio observations suggest that the large scale field lies
generally parallel to the disk, there may be small-scale variations in
the magnetic field geometries along which dust could be driven.  The
fact that vertical fields do not dominate the radio continuum
observations suggests, however, that coherent vertically-oriented
fields stretching for large distances are not likely; thus it would be
difficult to drive grains to the very large $z$-heights suggested by
the Ferrara et al. (1991) models.  However, for the structures
apparent in the WIYN images, it may be the case that the local fields
run perpendicular for long enough distances to allow radiation
pressure expulsion of grains to the requisite heights.

\subsection{Ejection via Dynamical Instabilities}

It may be possible to lift material from the thin disk via dynamical
instabilities.  Binney (1981) and Mulder \& Hooimeyer (1984) have
studied resonant excitation of $z$-directed instabilities in the
collisionless motions of stars, concluding that stars in flattened
potentials may be driven in the $z$-direction via resonant coupling of
$z$-oscillations and changes in the radial component of the force.
The coupling requires highly eccentric orbits or non-axisymmetric
potentials, i.e., bars.  For disk galaxies this coupling occurs in
only relatively thin annuli at specific radii.  More recently Erwin \&
Sparke (1997, in preparation) have investigated models of pre-main
sequence gaseous circumbinary accretion disks by following the orbits
of particles which have a fixed artificial viscosity; they similarly
find that material may be excited to highly inclined orbits via
dynamical instabilities and resonances.

All of these models rely upon resonances of particle orbits with
non-axisymmetric or bar-like potentials to excite the vertical
motions.  Though there may be evidence for a central bar in NGC 891
(Garc\'{\i}a-Burillo \& Gu\'{e}lin 1995), there still exists the
problem of energy dissipation in self- or disk-crossing orbits.  These
investigations have all studied {\em collisionless} orbits, while for
the present problem of gaseous material, orbits that are self-crossing
or disk plane-crossing may lead to significant energy loss and
possibly shocks.  It has yet to be seen whether gaseous material can
fully participate in these trajectories given its collisional nature.
It may be the case that parcels of the ISM can be thrown out of the
plane of a barred galaxy, but that the trajectory may be significantly
altered as it nears the plane by the material's interaction with disk
gas.  Also the orbits described by Binney (1981) and Mulder \&
Hooimeyer (1984) rely upon resonance effects which, for a
characteristic $z$-oscillation frequency, occur in thin annuli at
specific distances.  Thus if the dust seen away from the plane of NGC
891 were expelled via this type of resonant coupling, we would expect
to find no high-$z$ material beyond a cut-off point at some projected
distance from the center of the galaxy, which is not seen.

Dynamical instabilities as a whole can probably not be rejected at
this time, though it is not clear that they are capable of producing
the wide-spread extraplanar dust seen in NGC 891.

\subsection{Why NGC 891?}

It is not the case that all galaxies show an extensive network of
extraplanar dust structures like that seen in NGC 891.  A counter
example to this phenomenon is the Sb galaxy NGC 4565, also at a
distance of $\sim$10 Mpc; initial WIYN images of this galaxy at
0.75\arcsec\ resolution show a few very localized examples of dust
stretching away from the midplane to high $z$, but nothing like the
pervasive extraplanar dust found in NGC 891.

While it is too early to definitively declare causes for the
phenomenon of extraplanar dust in NGC 891, it is important to point
out that NGC 891 is not typical of most galaxies in various respects,
extraplanar dust structures possibly being one.  The radio continuum
and \Ha\ halos found in NGC 891 are not common.  Hummel et al. (1991),
in a search for extended radio continuum halos around edge-on
galaxies, found only 10\% of the galaxies show any extended-$z$
emission; only 5\% of these were thought to be good candidates for the
further study of extraplanar emission.

Further, it is becoming increasingly clear that extraplanar \Ha\
emitting material is not common among galaxies, particularly
relatively smoothly distributed, wide-spread emission as seen about
NGC 891 (Rand 1996; Pildis et al. 1994b; Dettmar 1992).  Several
galaxies show filamentary extensions (e.g., NGC 5775), while some
exhibit a patchy distribution of ionized material (e.g., NGC 4217).
The few galaxies that seem to show evidence for smoothly distributed
or wide-spread extraplanar DIG generally have relatively high
far-infrared luminosities and surface brightnesses, which may indicate
that the existence and appearance of extraplanar \Ha\ emission from
galaxies depends on the star formation rate or star formation rate per
unit area of the underlying disk (Rand 1996).  The extraplanar DIG in
NGC 891 is significantly brighter than that from most other galaxies.

The ionizing spectrum required to produce the observed emission line
ratios from the ionized halo of NGC 891 is likewise quite extreme,
though the only point of comparison currently available is the DIG of
the Milky Way.  The very deep spectrum of the extraplanar DIG obtained
by Rand (1997) suggests that nearly 70\% of the He in this diffuse
layer may be singly-ionized.  This conclusion is based upon the ratio
of the \ion{He}{1} 5876 \AA\ line to \Ha; the input spectrum required
to provide this level of ionization is significantly harder than that
required by the Reynolds \& Tufte (1995) result for DIG in the Milky
Way.

If the strength of the extraplanar DIG and radio continuum emission in
NGC 891 is indeed related to vigorous star formation in the underlying
disk, the spectacular extraplanar dust features seen in our WIYN
images may also be a result of the enhanced energy input associated
with high levels of star formation.  If these dust structures are the
result of a very high star formation rate within the disk of NGC 891,
they may be shaped by the hydrodynamical afterlives of massive stars
or the enhanced radiative driving force on dusty clouds due to the
greater number of these stars, or perhaps by the combination of the
two mechanisms.  We emphasize once again that the potential energies
derived in Sect. \ref{subsec:derived} are very much lower limits to
the input energies required to lift the observed dust structures from
the midplane.  In models involving explosively-driven expulsion, the
consideration of radiative losses and conversion of kinetic into
thermal energy may require as much as an order of magnitude greater
energy input in order to lift the clouds to their current positions
than those given in Table \ref{table:derived}.  Thus, if the
structures we see in NGC 891 are shaped primarily by the effects of
multiple supernovae, the energy input of several thousand such
explosions may be required to explain the $z$-heights of several of
these absorbing features.  This is a great deal of energy, though not
outside the realm of possibility for the larger OB associations.

Rand (1996) has found that several galaxies show quite localized
patches of extraplanar DIG (e.g., NGC 4217 and NGC 5023).  If these
cases of localized DIG are due to very locally enhanced star formation
rate, we might expect to find extraplanar dust filaments only in these
regions of the galaxies; for NGC 891, or other galaxies with
wide-spread \Ha\ emission, there may be high levels of star formation
activity throughout the disk, and thus also wide-spread extraplanar
dust features, if the dust structures are driven by star formation.
The dust features seen in NGC 891 may very well represent
extragalactic examples of the dusty ``worms'' of Koo et al. (1992),
particularly as seen in their IRAS 100 $\mu$m images, or the walls of
the ``chimneys'' postulated by Norman \& Ikeuchi (1989).  We note that
the star formation rate of NGC 4565, which does not show widespread
extraplanar dust features, is significantly lower than that of NGC 891
(as traced by the far-infrared luminosity, which is 20\% of that found
from NGC 891).  For now we can only speculate that the existence of
the dust structures discussed in this paper {\em may} be associated
with the relatively high star formation rate of NGC 891.

The discussion in Sect. \ref{subsec:thickdisk} regarding pure number
counts, while obviously incomplete, shows that $\sim 2 \times 10^8$
M$_\odot$ of gas may exist well beyond the thin disk of NGC 891.
Garc\'{\i}a-Burillo et al. (1992) comment that their extended
component of CO emission might have a total mass in the range $10^8 -
10^9$ M$_\odot$.  While this is not conclusive evidence, there may be
a connection between the dust seen in our WIYN images and the high-$z$
CO emission reported by Garc\'{\i}a-Burillo et al.  Also, the dusty
features seen here may be connected with high-velocity cloud
populations observed in spiral galaxies.  Schulman et al. (1994) find
several galaxies with high-velocity wings that may represent a
population of high-velocity clouds.  They derive gas masses for this
component in the range $6\times 10^7$ to $4 \times 10^9$ M$_\odot$,
4-14\% of the total estimated \HI\ mass of the galaxies in question.
The fractional and total masses of this material are not unlike our
estimates for the dust-containing features above the plane of NGC 891,
possibly also suggesting a connection between these components.

The masses of individual extraplanar dust features are similar to
those of the Galactic giant molecular clouds.  Might these dust
features be high-$z$ examples of such clouds?  If the masses derived
here are correct, or lower limits given our neglect of scattering, the
possibility exists that star formation may be occuring well away from
the plane of NGC 891 in some of these structures.  Particularly for
the most opaque structures, such as feature 1.  A careful search for
young extraplanar star clusters and possible associated \Ha\ emission
will be a goal for our next investigation of NGC 891.

\section{Summary}
\label{sec:summary}

  We present high-resolution (0.60-0.65\arcsec) optical BVR images of
the edge-on Sb galaxy NGC 891 obtained with the WIYN 3.5-m telescope.
The images reveal a wealth of halo absorbing dust structures viewed
against the background stellar light of the galaxy.  A study of these
structures leads us to the following conclusions:

1. The WIYN images reveal hundreds of dust absorption structures in
the disk-halo boundary of NGC 891.  The structures have a wide range
of often complex morphologies and are found up to $\sim$1.5 kpc from
the mid-plane of the galaxy.

2.  The dust features cannot be attributed to a warp because they
extend to both sides of the mid-plane of NGC 891. The features are not
likely to be associated with a flared gas layer at large
galactocentric distances because the WIYN data reveal evidence for
foreground stellar light and some of the features appear to be
associated with star formation events in the disk.  We attribute the
dust structures to inner galaxy (R$<$10 kpc) high-$z$ features.  The
dust features provide important information about the processes that
eject gas and dust into galaxy halos.

3. Twelve dust structures identified for detailed study are typically
hundreds of pc in length and 50 to 100 pc in width.  Assuming a Milky
Way hydrogen to dust extinction ratio, the inferred total hydrogen
column densities in these features range from $\sim$1 to
4$\times$10$^{21}$ cm$^{-2}$, and their masses range from $\sim$2 to
50$\times$10$^5$ M$_\odot$.  The gravitational potential energies of
the dust structures are greater than 20 to $\sim$200$\times$10$^{51}$
ergs.  If the high-$z$ dust is expelled from the disk, the input
energies may be significantly larger than the observed potential
energy.  These quantities should be regarded as {\em lower} limits.

4. The dust structures exhibit a wide range of complicated
morphologies. There are round and irregular clouds, vertical columns,
loop-like structures, cometary clouds and vertical cones.  The
higher-resolution HST image discussed herein shows the structures to
be filamentary on a scale even finer than that seen in the WIYN
images.  Some of the observed complexity may be due to overlapping
clouds along the line of sight.

5. We estimate at least 120 dust features exist at $| z | \gtrsim 400$
pc with $a_B \gtrsim 0.25$ mag.  Assuming Galactic gas to dust ratios,
the total implied mass of gas associated with these dust structures
exceeds $2 \times 10^8$ M$_\odot$, or approximately 2\% of the total
neutral gas mass of NGC 891.  There may be a significantly greater
number of structures than our counts suggest due to confusion along
the line of sight and our inability to pick out structures deep into
the galaxy because of foreground stellar light contamination.

6. Several dust structures are clearly associated with \Ha\ emission
structures that extend from regions of very active star formation in
the disk towards the halo.  The most pronounced examples are bubble-
and cone-shaped.

7.  The off-plane dust structures have apparent extinction properties
best characterized by $R_V \equiv A_V / E(B-V) \approx 3.6 \pm 0.4$,
where \Rv\ is the ratio of V-band selective to total extinction.  This
value is roughly consistent with the typical \Rv\ for Galactic diffuse
clouds.

8. The effects of large amounts of extinction, due to the presence of
extensive quantities of dust at large $z$-distances from the plane of
NGC 891, must be allowed for when attempting to interpret other data
relating to off-plane stellar and interstellar emission from NGC 891.

9. The high values of \Nh\ and large inferred masses for the more
prominent halo dust structures raises the possibility that much of the
hydrogen in these features is molecular and that some of the clouds
may be sites for current or future star formation.

10. We have reviewed a number of expulsion mechanisms that might
explain the existence of extraplanar gas and dust in the halos of
spiral galaxies.  In several cases the dust features we observe have
shapes and energy requirements suggestive of ejection phenomena driven
by multiple supernova, such as galactic chimney and fountain
processes.  Others have appearances open to a wide range of possible
explanations.

11. The processes that lift gas and dust into galaxy halos evidently
are not violent enough to completely destroy the dust grains.
Important non-destructive processes worthy of detailed study include
low-temperature galactic fountains, the radiative expulsion of grains
and various magnetic and dynamical processes.

\acknowledgements

JCH recognizes the guidance of Drs. G.D. Nickas, D.L. Steinert and
H.S. Thompson.  We also owe thanks to J. Gallagher, J. Mathis and
A. Ferrara for their comments and suggestions, as well as C. Corson,
D. Sawyer and T. von Hippel for their expert help with technical
issues.  We appreciate the help of A. Cole in dealing with the WFPC2
filter extinction characteristics.  Special thanks goes to
Dr. N. Sharp of NOAO for producing the true color image shown here as
Figure \ref{fig:WIYNcolor}.  We thank the anonymous referee for a
careful reading of this text and for making valuable suggestions on
its improvement.  Our deep appreciation is extended to the many people
who had significant roles in the planning and construction of the WIYN
Observatory and in the optimization of the scientific performance of
the telescope. We also acknowledge the major support from the higher
administration of the University of Wisconsin-Madison which made
possible our department's participation in the WIYN Observatory
project.  JCH and BDS acknowledge support from NASA grant NAG5-1852.

\pagebreak

\begin{table}
\dummytable\label{table:log}
\end{table}

\begin{table}
\dummytable\label{table:observed}
\end{table}

\begin{table}
\dummytable\label{table:derived}
\end{table}

\pagebreak

\figcaption{A three-color (BVR) image of NGC 891 as seen with the full
6.7\arcmin $\times$6.7\arcmin\ field of the WIYN CCD.  In this image
North is 20$^\circ$ clockwise from the top and East is similarly
angled from the left edge.  The seeing limited resolutions of the
individual BVR images are 0.60\arcsec -0.65\arcsec.  The resolution of
this three color composite image is slightly worse.  Note the presence
of large numbers of absorbing dust far from the mid-plane of the
galaxy.  (We thank Dr. N. Sharp of NOAO for processing this image.)
\label{fig:WIYNcolor}}

\figcaption{A B-band view of NGC 891; this figure has a resolution of
0.65\arcsec, slightly better than that of Figure \ref{fig:WIYNcolor}.
North and East are indicated on this figure; the bar in the lower left
corner of the image represents 1 kpc at a distance of 9.5 Mpc. Again,
numerous filamentary dust structures are seen far from the thin disk
of NGC 891, some at heights of at least 1.5 kpc from the
mid-plane. \label{fig:WIYNfull}}

\pagebreak

\figcaption{(a) A B-band close-up of the central bulge of NGC 891 with
0.65\arcsec\ seeing and (b) an unsharp-masked version of this same
region.  The region displayed covers 2.2\arcmin $\times$1.5\arcmin.
The scale of the image is given in kpc on the side and bottom of the
figure, referring to the $z$-height and projected radial distance from
the center, respectively.  Some stars have been removed in producing
the unsharp-masked image.  \label{fig:WIYNcenter} }

\figcaption{As Figure \ref{fig:WIYNcenter}(a) and (b), but for a
section of the disk to the NE of the bulge.  This section is $\sim$5
kpc from the center of the galaxy.  This is a region of very active
star-formation and shows dust structures extending to high-$z$ that
are likely caused by the effects of multiple
supernovae. \label{fig:WIYNne}}

\figcaption{As Figure \ref{fig:WIYNcenter}(a) and (b), but for a
section of the disk to the SW of the bulge. This section is $\sim$5
kpc from the center of the galaxy and seems to have a significantly
lower star-formation rate than the region shown in Figure
\ref{fig:WIYNne}.  Several stars that lay directly on the disk or in
the low halo that have been removed can be seen in the unsharp-masked
version of this image.
\label{fig:WIYNsw}}

\pagebreak

\figcaption{Apparent extinction for B- and R-bands
(squares and triangles, respectively) versus that in V-band for
high-$z$ dust in NGC 891.  The solid lines represent the Galactic
relationship which would hold if the dust structures were all on the
front side of the galaxy with \Rv=3.6.  The dotted lines are the
values expected for the apparent extinction in the B- and R-bands if
the clouds have a fraction $x$ of the stellar light in front of them.
The numbers on the plot give the values of $x$.  Also plotted as
dashed lines are the expected behaviors for clouds characterized by
\Rv=3.2 and 4.0 (outer and inner dashed lines, respectively).  These
values are the 1 $\sigma$ deviations from the best fit to the data.
The cross in the lower right shows the approximate $\pm1\sigma$ error
bars for the data points.
\label{fig:extinct} }

\figcaption{(a) A B-band close-up of the section to the NE of the
bulge [same as \ref{fig:WIYNne}(a)] and (b) an image of ionized gas in
this region (\Ha+\NII). A comparison of the B-band and ionized gas
images show correspondences between structures seen in dust absorption
and ionized gas. It should be noted that this image is inverted across
the horizontal axis from that shown in Pildis et al. (1994).
\label{fig:halpha}}

\figcaption{The 180-s F606W WFPC2 image of the central region of NGC
891 at 0.10\arcsec\ resolution.  This figure should be compared with
the WIYN image of a similar region given in Figure
\ref{fig:WIYNcenter}, though the WIYN image shown in that figure was
obtained in the B-band. \label{fig:wfpc2} }

\pagebreak

\figcaption{Close-ups of features 1 and 2 as seen by WIYN and HST
(left and right, respectively).  The WIYN close-up is from the B-band
image, while the HST image was taken through the F606W filter.  The
region displayed is 30\arcsec, or $\approx$1.3 kpc, on a side.
Though the B-band image is more sensitive to low apparent extinction
features, the differences in resolution are shown clearly.
\label{fig:feat1}}

\figcaption{As for Figure \ref{fig:feat1}, but centered on feature 4,
also showing features 3. Again, the region displayed is 30\arcsec, or
$\approx$1.3 kpc, on a side.  \label{fig:feat4}}

\end{document}